\documentclass[12pt]{article}

\usepackage{scicite}
\usepackage{graphicx}

\newenvironment{sciabstract}{%
\begin{quote} \bf}
{\end{quote}}

\newcounter{lastnote}

\title{The metamorphosis of Supernova SN\,2008D/XRF\,080109: a link between
Supernovae and GRBs/Hypernovae\\[3.5cm] 
{\Large Accepted by Science}}

\author{Paolo~A.~Mazzali$^{1,2,3,4\ast}$,
Stefano Valenti$^{5,6}$,
Massimo Della Valle $^{7,8,9}$,\\
Guido Chincarini$^{10,11}$,
Daniel N. Sauer$^{1}$,
Stefano Benetti$^{2}$,
Elena Pian$^{12}$,\\
Tsvi Piran$^{13}$,
Valerio D'Elia$^{14}$,
Nancy Elias-Rosa$^{1}$,
Raffaella Margutti$^{10}$,\\
Francesco Pasotti$^{10}$,
L. Angelo Antonelli$^{14}$,
Filomena Bufano$^{2}$,\\
Sergio Campana$^{11}$,
Enrico Cappellaro$^{2}$,
Stefano Covino$^{11}$,\\
Paolo D'Avanzo$^{11}$,
Fabrizio Fiore$^{14}$,
Dino Fugazza$^{11}$,
Roberto Gilmozzi$^{8}$,\\
Deborah Hunter$^{5}$,
Kate Maguire$^{5}$,
Elisabetta Maiorano$^{15}$,
Paola Marziani$^{2}$,\\
Nicola Masetti$^{15}$,
Felix Mirabel$^{16}$,
Hripsime Navasardyan$^{2}$,\\
Ken'ichi Nomoto$^{3,4,17}$,
Eliana Palazzi$^{15}$,
Andrea Pastorello$^{5}$,\\
Nino Panagia$^{18,19}$,
L.J. Pellizza$^{20}$,
Re'em Sari$^{13}$,
Stephen Smartt$^{5}$,\\
Gianpiero Tagliaferri$^{11}$,
Masaomi Tanaka$^{3}$,
Stefan Taubenberger$^{1}$,\\
Nozomu Tominaga$^{3}$,
Carrie Trundle$^{5}$,
Massimo Turatto$^{19}$}

\date{}

\newcommand{\eg}{e.g.\ }

\newcommand{\Msun}{M_{\odot}}
\newcommand{\kms}{km\,s$^{-1}$}

\newcommand{\HII}{H~{\sc ii}}
\newcommand{\CI}{C\,{\sc i}}
\newcommand{\OI}{O\,{\sc i}}

\newcommand{\OIII}{O~{\sc iii}}
\newcommand{\HeI}{He\,{\sc i}}

\newcommand{\NaI}{Na\,{\sc i}}
\newcommand{\SiII}{Si\,{\sc ii}}
\newcommand{\CaII}{Ca\,{\sc ii}}
\newcommand{\FeII}{Fe\,{\sc ii}}

\newcommand{\Fefs}{$^{56}$Fe}
\newcommand{\Cofs}{$^{56}$Co}
\newcommand{\Nifs}{$^{56}$Ni}
\newcommand{\Mej}{$M_{\rm ej}$}
\newcommand{\KE}{$E$}
\newcommand{\msun}{\ensuremath{M_{\odot}}}

\def\gsim{\mathrel{\rlap{\lower 4pt \hbox{\hskip 1pt $\sim$}}\raise 1pt \hbox {$>$}}}
\def\lsim{\mathrel{\rlap{\lower 4pt \hbox{\hskip 1pt $\sim$}}\raise 1pt \hbox {$<$}}}

\def\simlt{\mathrel{\hbox{\rlap{\hbox{\lower4pt\hbox{$\sim$}}}\hbox{$<$}}}}
\def\simgt{\mathrel{\hbox{\rlap{\hbox{\lower4pt\hbox{$\sim$}}}\hbox{$>$}}}}

\newcommand{\ergsec}{\mbox{erg s$^{-1}$}}

\newcommand{\ergcms}{\mbox{erg cm$^{-2}$ s$^{-1}$}}

\def\apj{ApJ}%
\def\apjl{ApJ}%
\def\apjs{ApJS}%
\def\aap{A\&A}%
\def\mnras{MNRAS}%
\def\pasp{PASP}%
\def\nat{Nature}%
\begin{document}

% Double-space the manuscript.

%\renewcommand{\baselinestretch}{2}
%\baselineskip24pt

% Make the title.

\maketitle
\clearpage
\noindent
\normalsize{$^{1}$Max-Planck Institut f\"ur Astrophysik,
    Karl-Schwarzschild-Str.1, 85748 Garching, Germany}\\
\normalsize{$^{2}$Istituto Nazionale di Astrofisica-OAPd, vicolo dell'Osservatorio, 2, I-35122 Padova, Italy}\\
\normalsize{$^{3}$Department of Astronomy, School of Science, University of Tokyo, Bunkyo-ku, Tokyo 113-0033, Japan}\\
\normalsize{$^{4}$Research Center for the Early Universe, School of Science, University of Tokyo, Bunkyo-ku, Tokyo 113-0033, Japan}\\
\normalsize{$^{5}$Astrophysics Research Centre, School of Maths and Physics, Queen's University, Belfast, BT7 1NN, Northern Ireland, UK}\\
\normalsize{$^{6}$Dipartimento di Fisica, Universita' di Ferrara, via G. Saragat 1, 44100 Ferrara, Italy}\\
\normalsize{$^{7}$Istituto Nazionale di Astrofisica, Capodimonte Astronomical Observatory, Salita Moiariello 16, I-80131 Napoli, Italy}\\
\normalsize{$^{8}$European Southern Observatory, Karl-Schwarzschild-Str.2, D-85748 Garching, Germany}\\
\normalsize{$^{9}$International Center for Relativistic Astrophysics Network, Piazzale della Repubblica 10, I-65122 Pescara, Italy}\\
\normalsize{$^{10}$Department of Physics, Universita' di Milano-Bicocca, Piazza delle Scienze 3, I-20126 Milano, Italy}\\
\normalsize{$^{11}$Istituto Nazionale di Astrofisica, Brera Astronomical Observatory, Via E. Bianchi 46, I-23807 Merate (LC), Italy}\\
\normalsize{$^{12}$Istituto Nazionale di Astrofisica-OATs, Via Tiepolo 11, I-34131 Trieste, Italy}\\
\normalsize{$^{13}$The Racah Institute of Physics, Hebrew University, Jerusalem 91904, Israel}\\
\normalsize{$^{14}$Istituto Nazionale di Astrofisica, Rome Astronomical Observatory, Via di Frascati 33, I-00040 Monte Porzio Catone, Italy}\\
\normalsize{$^{15}$Istituto Nazionale di Astrofisica, Istituto di Astrofisica Spaziale e Fisica Cosmica, Via P. Gobetti 101, I-40129 Bologna, Italy}\\
\normalsize{$^{16}$European Southern Observatory, Alonso de Cordova 3107, Santiago, Chile}\\
\normalsize{$^{17}$Institute for the Physics and Mathematics of the Universe, University of Tokyo, Kashiwa, Chiba 277-8582, Japan}\\
\normalsize{$^{18}$Space Telescope Science Institute, Baltimore, MD, USA}\\
\normalsize{$^{19}$Istituto Nazionale di Astrofisica, Catania Astronomical Observatory, Via S. Sofia 78, I-95123 Catania, Italy}\\
\normalsize{$^{20}$Instituto de Astronom\'{\i}a y F\'{\i}sica del Espacio, C.C. 67, Buenos Aires, Argentina}\\
\\
\normalsize{$^\ast$To whom correspondence should be addressed; E-mail:
mazzali@mpa-garching.mpg.de}\\
\clearpage
% Place your abstract within the special {sciabstract} environment.

\begin{sciabstract}

The only supernovae (SNe) to have shown early $\gamma$-ray or X-ray emission
thus far are overenergetic, broad-lined Type Ic SNe (Hypernovae - HNe).
Recently, SN\,2008D shows several novel features: (i) weak XRF, (ii) an early,
narrow optical peak, (iii) disappearance of the broad lines typical of SN\,Ic
HNe, (iv) development of He lines as in SNe\,Ib.  Detailed analysis shows that
SN\,2008D was not a normal SN: its explosion energy ($\KE \approx 6 \cdot
10^{51}$\,erg) and ejected mass ($\sim 7 \Msun$) are intermediate between normal
SNe\,Ibc and HNe. We derive that SN\,2008D was originally a $\sim 30 \Msun$
star. When it collapsed a black hole formed and a weak, mildly relativistic jet
was produced, which caused the XRF. SN\,2008D is probably among the weakest
explosions that produce relativistic jets. Inner engine activity appears to be
present whenever massive stars collapse to black holes.

\end{sciabstract}

%%%%%%%%%%%%%%%%%%%%%%%%%%%%%%%%%%%%%%%%%%%%%%%%%%%%%%%%%%%%%%%%%%%

On 2008 January 9.57\,UT the X-Ray Telescope (XRT) on board Swift detected a
weak X-ray Flash (XRF\,080109) in the galaxy NGC2770(1).  Optical follow-up
revealed the presence of a supernova coincident with the XRF [SN\,2008D;
RA(2000) = 09 09 30.625; Dec (2000) = +33 08 20.16](2).  We detected SN\,2008D
photometrically from Asiago Observatory on 10.01 January 2008\,UT, only 10.5
hours after the Swift detection. Early spectra showed broad absorption lines
superposed on a blue continuum, and lacked hydrogen or helium lines(3).
Accordingly, SN\,2008D was classified as a broad-lined SN\,Ic(4). SNe of this
type are sometimes associated with Gamma-ray Bursts (GRB, Refs. 5,6) or XRFs
(7,8).  The spectra resembled those of the XRF-SN\,2006aj (8) or the non-GRB HN
SN\,2002ap (9) (Fig. 3, top), but a comparison suggests that SN\,2008D was
highly reddened: we estimate that $E(B-V)_{tot} = 0.65$ mag (see SOM).

The host galaxy of XRF\,080109/SN\,2008D, NGC\,2770 [redshift $z=0.006494$, distance
31 Mpc], is a spiral galaxy similar to the Milky Way, M31, or ESO\,184-G82, the
host of SN\,1998bw/GRB\,980425. NGC\,2770 has roughly solar metallicity and a
moderate star-formation rate, $\sim 0.5 \Msun$\,yr$^{-1}$ (see SOM).  In
contrast, typical host galaxies of GRBs are small, compact, somewhat metal-poor,
and highly star-forming(10).

In addition to the weak XRF, SN\,2008D shows a number of peculiar features, most
of which are new. The optical light curve had two peaks (Fig. 1): a first, dim
maximum ($V \approx 18.4$) was reached less than 2 days after the XRF. After a
brief decline the luminosity increased again, reaching principal maximum ($V =
17.37$) $\sim 19$ days after the XRF.  An 18-20 day risetime is typical of
GRB-HNe: normal SNe\,Ic reach maximum in 10-12 days. Few stripped-envelope SNe
have very early data, and in GRB-HNe a first peak may be masked by the afterglow
light. A first narrow optical peak was only seen in the  Type Ib SN\,1999ex
(SNe\,Ib are similar to SNe\,Ic but show strong helium lines, Ref. 4), the Type
IIb SN\,1993J (SNe\,IIb are similar to SNe\,Ib but still have some hydrogen),
and the Type Ic  XRF-SN\,2006aj. When it was discovered, SN\,1999ex was dropping
from a phase of high luminosity(11). It reached principal maximum $\sim 20$ days
later, as did SN\,2008D.

Another novel feature is the spectral metamorphosis (Fig. 2). Unlike SNe\,2006aj
and 2002ap, the broad absorptions did not persist. As they disappeared, \HeI\
lines developed (12). By principal maximum SN\,2008D had a narrow-lined, Type Ib
spectrum (Fig. 3, bottom).

Broad lines require material moving with velocity $v > 0.1c$, where $c$ is the
speed of light (13). Their disappearance implies that the mass moving at  high
velocities was small.

Late development of \HeI\ lines, previously seen only in SN\,2005bf (14), is
predicted by theory (15). Helium levels have high excitation potentials,
exceeding the energy of thermal photons and electrons.  Excitation can be
provided by the fast particles produced as the $\gamma$-rays emitted in the
decay chain of \Nifs\ thermalize (16). This is the process that makes SNe shine.
In the first few days after explosion thermalization is efficient because of the
high densities and not enough particles are available to excite helium. Only
when density drops sufficiently can more particles escape the \Nifs\ zone and
excite helium.

We reproduced the spectral evolution and the light curve of SN\,2008D after the
first narrow peak using a model with \Mej\,$\sim 7 \Msun$ and spherically
symmetric \KE\,$\sim 6 \cdot 10^{51}$\,erg, of which $\sim 0.03 \Msun$, with
energy $\sim 5 \cdot 10^{50}$\,erg, are at $v > 0.1c$ (Figs. 1, 3, and SOM). Our
light curve fits indicate that SN\,2008D synthesised $\sim 0.09 \Msun$ of \Nifs,
like the non-GRB HN SN\,Ic 2002ap (9) and the normal SN\,Ic 1994I (17)  but much
less than the luminous GRB-HN SN\,1998bw (6). The rapid rise in luminosity
following the first peak requires that some \Nifs\ $(0.02 \Msun)$ was mixed
uniformly at all velocities $> 9000\,$\kms. This is a typical feature of HNe,
and indicates an aspherical explosion (18). Asphericity may affect our estimate
of the energy, but not the \Nifs\ mass (19).

Comparing the mass of the exploding He-star that we derived with evolutionary
models of massive stars, we find that the progenitor had main sequence mass
$\sim 30 \Msun$. A star of this mass is likely to collapse to a black hole, as
do GRB/SNe (20). So, SN\,2008D shared several features of GRB/HNe. However, all
SNe with GRBs or strong XRFs initially had velocities higher than SN\,2008D or
SN\,2002ap (Fig.\ S3) and never showed helium. Had the He layer not been present
in SN\,2008D, the explosion energy would have accelerated the inner core to
higher velocities, and broad lines may have survived.

The characterizing features of SN\,2008D (weak XRF, first narrow optical peak,
initially broad-lined SN\,Ic spectrum that later transformed into a narrow-lined
SN\,Ib spectrum) may be common to all SNe\,Ib, or at least a significant
fraction of them, and maybe some SNe\,Ic, which however contain little or no
helium. The light curves of various SNe\,Ib are rather similar (21).  The first
peak was observed only for SN\,1999ex, but lack of X-ray monitoring probably
prevented the detection of more weak XRFs and the early discovery of the
associated SNe. On the other hand SN\,2008D, and possibly most SNe\,Ib, was more
energetic than normal core-collapse SNe, including most SNe\,Ic.

Type II SNe in late Spiral/Irr galaxies (the typical Hubble type of GRB hosts)
are about 6 times more frequent than SNe\,Ib (22). Although the serendipitous
discovery of an SN\,Ib by XRT may be a statistical fluctuation, it may also
suggest that the soft X-ray emission accompanying SN\,2008D is typical of
overenergetic SNe\,Ib, and absent (or very weak) in normal core-collapse SNe.

The X-ray spectrum of SN\,2008D (in total $\sim 500$ photons) can be fitted with
either a simple power-law indicating a non-thermal emission mechanism or a
combination of a hot black body $(T = 3.8 \cdot 10^6$\,K) and a power law.  In
the latter case, the unabsorbed luminosity of the black-body component is a
small fraction of the total X-ray luminosity. The high temperature and low
luminosity ($L = 1.1 \cdot 10^{43}$\,erg\,s$^{-1}$) of the black-body component
at first peak ($\sim 100$\,s after the onset of the XRF) imply an emitting
radius $R_{{\rm ph}} \sim 10^{10}$\,cm (see SOM, Section 4). This is at least
one order of magnitude smaller than the size of Wolf-Rayet stars, the likely
progenitors of SNe\,Ibc. 

The X-ray flare and the first optical peak are most likely associated (23). The
timescale of the first optical peak may suggest that it was related to shock
breakout.  A signature of shock breakout is a hot black-body X-ray spectrum
immediately after the explosion. Thermal X-ray emission was suggested for
SN\,2006aj (24), while no X-ray data are available for SN\,1999ex.  The model of
Ref.\ 23 uses a spherical configuration and a black-body component at
$\sim$0.1\,keV, below the XRT energy range. This yelds a large radius, which
they explain invoking the presence of a dense surrounding medium that
bulk-Comptonize the shock breakout emission to higher energies, producing the
power-law spectrum observed by XRT between 0.3 and 10\,keV.

On the other hand, the angular size of an emitting area with radius $R_{{\rm
ph}} \sim 10^{10}$\,cm  is typical of GRB jets. This leads naturally to an
alternative scenario, that we propose here:  XRF\,080109 was the breakout of a
failed relativistic jet powered by a central engine as in GRBs. The jet failed
because its energy was initially low or because it was damped by the He layer,
which is absent in GRB-HNe, or both. The presence of a jet is supported by our
conclusion that a black hole was probably formed when the star collapsed. The
marginal breakout of the jet produced thermal X-rays and relativistic particles
that caused the power-law X-ray component.  It also caused the first optical
peak: the timescale of the first peak and the X-ray flare and the corresponding
radii and temperature are consistent with emission from rapidly expanding,
adiabatically cooling material. The weakness of the jet resulted in the low
X-ray flux and the small amount of material with $v > 0.1c$. The failed jet
contributed anisotropically to the SN kinetic energy. Lateral spreading of the
ejecta with $v > 0.1c$ leads to an angular size larger than the X-ray-emitting
region, which is needed to produce the observed broad lines. The small amount of
high-velocity material moving along our line-of-sight may also indicate that we
viewed the explosion significantly off-axis. This can be tested by polarization
or line profiles studies at late times, as in SN\,2003jd (25).  The jet will
spread further after breakout and it could dominate the radio emission at later
times.

The scenario we propose implies that GRB-like inner engine activity exists in
all black hole-forming SNe\,Ibc (26).  SN\,2008D (and probably other SNe\,Ib)
has significantly higher energy than normal core-collapse SNe, although less
than GRB/HNe. Therefore, it is unlikely that all SNe Ibc, and even more so all
core-collapse SNe produce a weak X-ray flash similar to XRF\,080109. The
presence of high-energy emission (GRB, XRF) depends on the jet energy and the
stellar properties. Only massive, energetic, stripped SNe\,Ic (HNe) have shown
GRBs. In borderline events like SN\,2008D only a weak, mildly relativistic jet
may emerge, because the collapsing mass is too small and a He layer damps the
jet. For even less massive stars that still collapse to a black hole producing a
less energetic explosion (\eg SN\,2002ap) no jet may emerge at all. Stars that
only collapse to a neutron star are not expected to have jets. SN\,2008D thus
links events that are physically related but have different observational
properties.

%\noindent
%Correspondence should be addressed to P.A.\,Mazzali
%(e-mail: mazzali@MPA-Garching.MPG.DE).

%\begin{acknowledge}
%\end{acknowledge}

\renewcommand{\baselinestretch}{1}

%%%%%%%%%%%%%%%%%%%%%%%%%%%%%%% Figure 1 %%%%%%%%%%%%%%%%%%%%%%%%%%%%%%%%%%%%%%%%

\begin{figure}
%\centerline{\psfig{file=Fig1_sneLCbol.eps,width=16.cm}}
\vspace{-3cm}
\centerline{\includegraphics[width=16.cm]{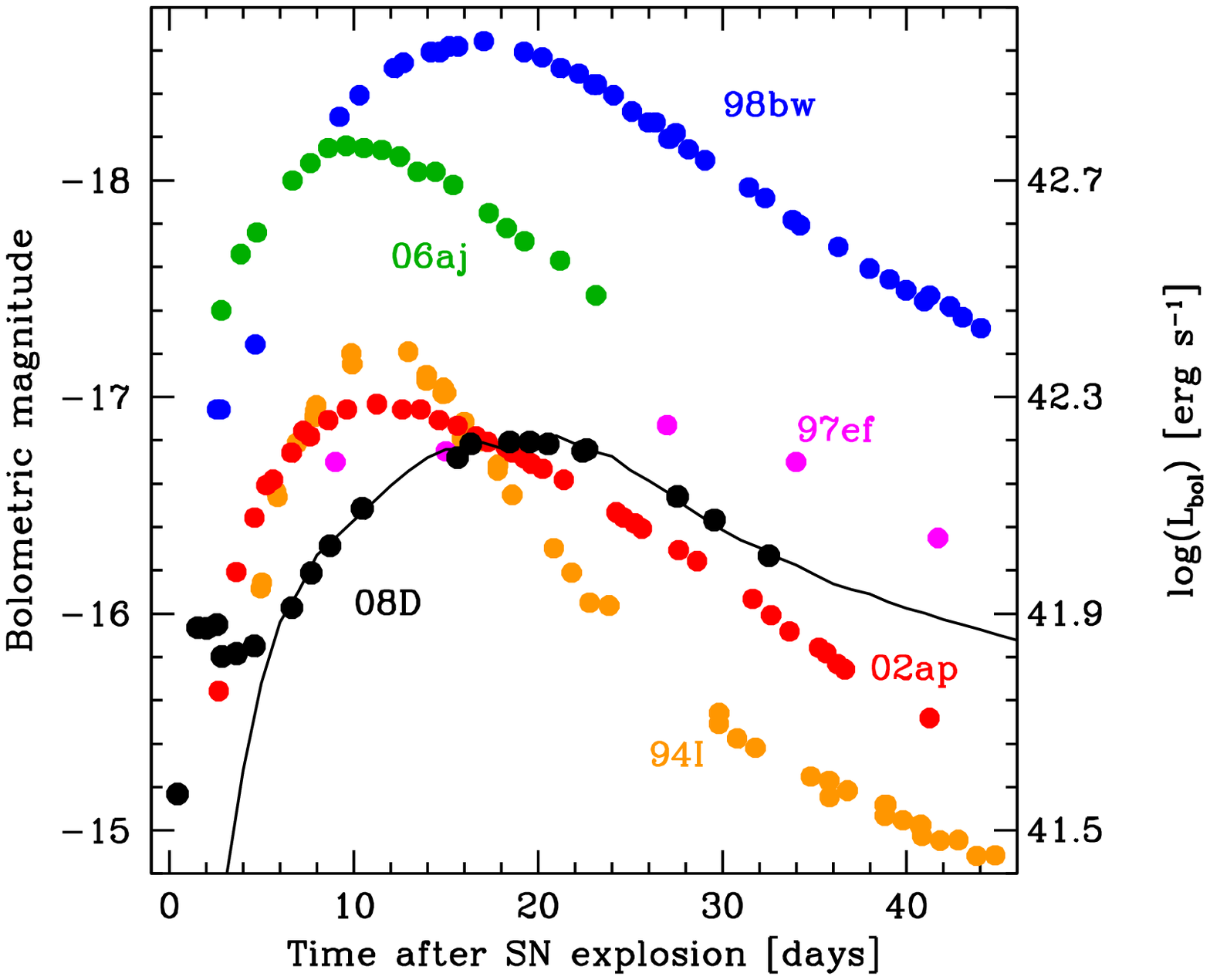}}
\caption{\footnotesize
The light curves of SN\,2008D and of other Type Ibc SNe.
The shape of the light curve of SN\,2008D is similar to that of SN\,1998bw
and other GRB/HNe, and comparable to the non-GRB HN SN\,1997ef, but much
broader than the XRF/SN 2006aj or the normal SN\,Ic 1994I. This similarity
suggests a comparable value of the quantity \Mej$^3 / $\KE, where \Mej\ is the
mass ejected and \KE\ the explosion kinetic energy(27). All known SNe\,Ic with a
broad light curve ejected a large mass of material(28). Large values of \Mej\
and \KE\ are also suggested by the presence of He moving at $v \sim
10000$\,\kms: the velocity of He in SN\,2005bf was lower(14). The light curve of
SN\,1999ex, which is similar to that of SN\,2008D, was fitted reasonably well by
a He-star explosion model with \Mej$\sim 5 \Msun$, \KE$\sim 3 \cdot
10^{51}$\,erg(11). Such a model would also match the light curve of SN\,2008D,
but it probably would not reproduce the broad lines that characterize the early
spectra. This would require a model containing some high-velocity material,
leading to a larger \KE\ without significantly affecting the value of \Mej\ or
the light curve shape. The line shows a synthetic bolometric light curve
computed with a Montecarlo code(29) for a model with \Mej$\sim 7 \Msun$,
\KE$\sim 6 \cdot 10^{51}$\,erg. The model does not address the physics that may
be responsible for the first narrow light curve peak, but only the main peak,
which is due to diffusion of radiation in the SN envelope following the
deposition of $\gamma$-rays and positrons emitted in the decay chain \Nifs\ to
\Cofs\ and \Fefs.
\label{fig:fig1}}
\end{figure}

\clearpage

%%%%%%%%%%%%%%%%%%%%%%%%%%%%%%%% Figure 2 %%%%%%%%%%%%%%%%%%%%%%%%%%%%%%%%%%%%%%

\begin{figure}
%\centerline{\psfig{file=Fig2_specevol.eps,width=14.cm}}
\vspace{-3cm}
\centerline{\includegraphics[width=14.cm]{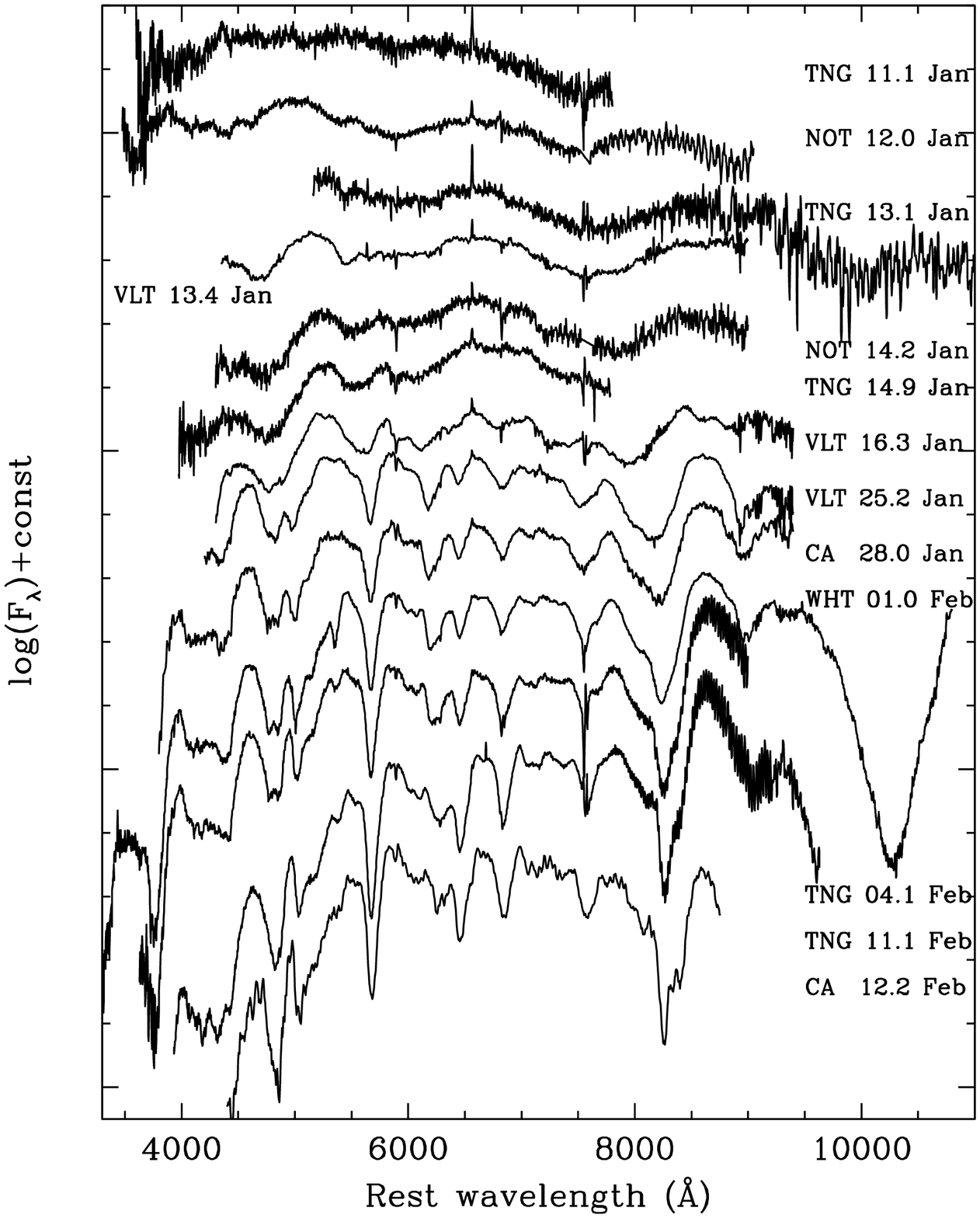}}
\caption{\footnotesize
Spectral evolution of SN\,2008D.
In the early phase the strongest features are broad Fe complexes in the blue
($\sim 4000 - 5000$\,\AA), the \SiII-dominated feature near 6000\AA, and \CaII\
lines, both in the near UV (H\&K) and in the near IR (the IR triplet near
8500\AA). On the other hand \OI\,7774\,\AA, which is strong in all HNe as well
as in all SNe Ibc, is conspicuously missing. Starting 15 Jan, lines begin to
become narrow.
In the later spectra, taken near maximum, \HeI\ lines have developed. The
strongest isolated lines are $\lambda 6678$\,\AA, seen near 6500\AA, and
$\lambda 7065$\,\AA, seen near 6900\AA. Both lines indicate a helium velocity
of $\sim 10000$\,\kms. The other strong optical lines of \HeI\ are blended:
$\lambda 5876$\,\AA\ is blended with \NaI\,D, near 5600\,\AA, and
$\lambda 4471$\,\AA\ is blended with the broad \FeII\ trough near 4200\,\AA.
\label{fig:fig2}}
\end{figure}

\clearpage

%%%%%%%%%%%%%%%%%%%%%%%%%%%%%%%%% Figure 3 %%%%%%%%%%%%%%%%%%%%%%%%%%%%%%%%%%%%%

\begin{figure}
%\centerline{\psfig{file=Fig3_speccomp.eps,width=16.cm,angle=270}}
\centerline{\includegraphics[width=11.cm,angle=270]{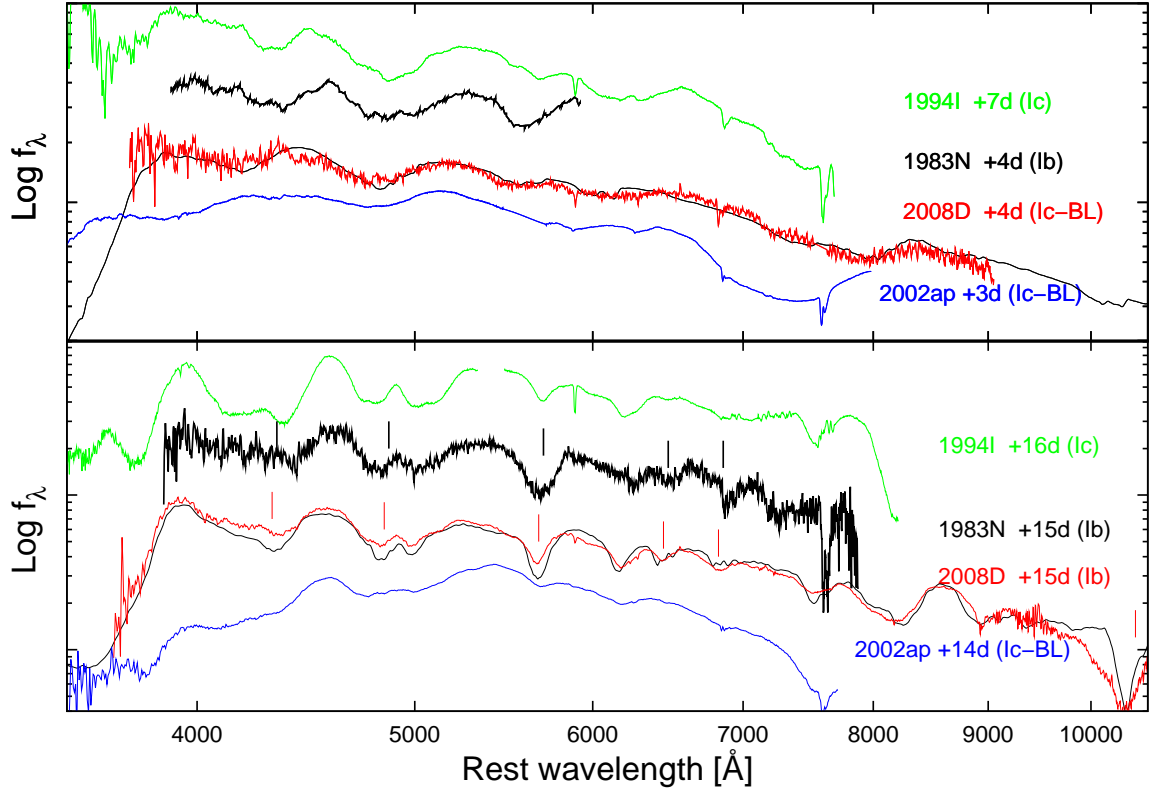}}
\caption{\footnotesize The spectra of SN\,2008D compared to those of
other Type Ibc SNe and to simulations. Near the first peak (top),
SN\,2008D has a broad-lined spectrum similar to that of SN\,2002ap,
a broad-lined SN without a GRB(9), but different from both the
normal SN\,Ic 1994I and the SN\,Ib 1983N. At the time of the main
light curve peak (bottom), the spectrum of SN\,2008D has narrow
lines like SNe\,1994I and 1983N, while SN\,2002ap and other HNe
retain broad features throughout their evolution. Also, SN\,2008D
developed He lines (vertical ticks). At this epoch, the spectra of
SNe\,2008D, 1983N, and 1999ex are similar. Synthetic spectra
are overlaid on the two SN\,2008D spectra (see SOM).
\label{fig:fig3}}
\end{figure}

\clearpage

%%%%%%%%%%%%%%%%%%%%%%%%%%%%%%%%%%%%%%%%%%%%%%%%%%%%%%%%%%%%%%%%%%%%%%%%%%%%%%%%

\pagebreak
%   ------------------    SUPPLEMENTARY MATERIAL    --------------------

{\bf {\LARGE 
	Supplementary Information 
	\vspace{10mm} }}
%\date{}{}
%\maketitle

% ----------------------  1: OBSERVATIONS ---------------------
{\bf 1~~~Optical and infrared observations}

\bigskip

SN\,2008D was observed photometrically and spectroscopically with a number of
telescopes. 

Low-resolution spectra of SN\,2008D have been obtained with the European
Southern Observatory (ESO) 8.2m Very Large Telescope (VLT)-UT2, equipped with
FORS2; the 4.2m William Herschel Telescope (WHT) equipped  with ISIS; the 3.6m
Telescopio Nazionale Galileo (TNG), equipped with DOLORES for optical
observations and with NICS for near-IR observations; the 3.6m ESO-New Technology
Telescope (NTT), equipped with EMMI; the 2.5m Nordic Optical Telescope (NOT),
equipped with ALFOSC; the 2.2m Calar Alto telescope (CA), equipped with CAFOS;
the 2.0m Liverpool Telescope (LT), equipped with RATCam; the 1.82m Asiago-Ekar
telescope (As1.82m), equipped with AFOSC; the 0.60m Rapid Eye Mount telescope
(REM), equipped with ROSS for optical observations and with REMIR for near-IR
observations. 

Tables S1 and S2 report the log of the observations covering the first month
after the XRF, with details on the acquisition, and the photometric results. 
Column 1 gives the observation date, Column 2 the telescope and instrument used,
Column 3 the observing setup, Column 4 the average seeing during the photometric
acquisition (Table S1) and the spectral resolution (Table S2), Column 5 the
observed V magnitude (Table S1) and the spectrophotometric standard used to
calibrate the spectra (Table S2).  The photometric errors are 1$\sigma$
uncertainties.

Monochromatic light curves are shown in Fig. S1, and the spectral evolution in
{Fig. 2} in the main paper).

\clearpage

%%%%%%%%%%%%%%%%%%%%%%%%%%%%%%%%%%%%%%%%%%%%%%%%%%%%%%%%%%%%%%%%%%%%%%%%%%%%%%

\begin{table}
\begin{center}
\setlength{\tabcolsep}{4pt}
%\resizebox{!}{5cm}{\begin{tabular}{cccccc}
\begin{tabular}{ccccc}
\multicolumn{5}{c}{\bf Supplementary Table S1: 
Photometric observations of SN~2008D}\\
\hline
\hline
Date       & Telescope+          & Setup        & Seeing   & V  \\
(2008 UT)  & Instrument          &              & (arcsec) &     \\
\hline
 7.005  Jan  & As1.82m+AFOSC     & UBVRI        & 2.0      & $ < 19.0$ \\
10.012  Jan  & As1.82m+AFOSC     & UBVRI        & 2.5      & $ 19.10 \pm 0.06$ \\
11.213  Jan  & TNG+NICS          & JHK          & 0.9      &  --             \\
12.168  Jan  & NOT+ALFOSC        & UBVRI        & 2.0      & $ 18.41 \pm 0.10$ \\
13.006  Jan  & TNG+NICS          & JHK          & 0.8      &  --             \\
13.212  Jan  & TNG+DOLORES       & UBVRI        & 0.9      & $ 18.49 \pm 0.05 $ \\
14.174  Jan  & NOT+ALFOSC        & UBVRI        & 1.6      & $ 18.29 \pm 0.05 $ \\
16.205  Jan  & LT+RATCam         & UBVRI        & 1.9      & $ 18.03 \pm 0.04$\\ 
17.270  Jan  & LT+RATCam         & UBVRI        & 1.6      & $ 17.96 \pm 0.02$\\ 
18.267  Jan  & LT+RATCam         & UBVRI        & 1.9      & $ 17.79 \pm 0.04$\\
20.024  Jan  & LT+RATCam         & UBVRI        & 2.2      & $ 17.64 \pm 0.06$\\
20.153  Jan  & REM+REMIR         & H            & 3.5      & --              \\
21.310  Jan  & REM+REMIR         & H            & 2.9      & --              \\
22.309  Jan  & REM+REMIR         & H            & 2.9      & --              \\
23.123  Jan  & REM+REMIR         & JK           & 1.4      & --              \\
25.197  Jan  & VLT+FORS2         &  BVRI        & 1.0      & $ 17.36 \pm 0.04$\\ 
25.906  Jan  & LT+RATCam         & UBVRI        &  1.9     & $ 17.36 \pm 0.03$ \\
28.021  Jan  & CA+CAFOSC         & UBVRI        &  1.2     & $ 17.33 \pm 0.05$\\
28.987  Jan  & TNG+NICS          & JHK          &  1.4     & --              \\
29.128  Jan  & LT+RATCam         & UBVRI        &  1.2     & $ 17.34\pm 0.03$ \\
30.127  Jan  & LT+RATCam         & UBVRI        &   2.9    & $ 17.35 \pm 0.04 $  \\
31.239  Jan  & REM+ROS+REMIR     & RJHK         &   3.0    & --              \\ 
31.947  Jan  & TNG+NICS          & JHK          &   3.1    & --              \\
31.964  Jan  & LT+RATCam         & UBVRI        &   1.3    & $ 17.38\pm 0.05$\\
\hline
\end{tabular}
\end{center}
\end{table}
%%%%%%%%%%%%%%%%%%%%%%%%%%%%%%%%%%%%%%%%%%%%%%%%%%%%%%%%%%%%%%%%%%%%%%%%%%%%%%

\begin{table}[t]
\begin{center}
\setlength{\tabcolsep}{4pt}
%\resizebox{!}{5cm}{\begin{tabular}{cccccc}
\begin{tabular}{ccccc}
\multicolumn{5}{c}{\bf Supplementary Table S1: 
Photometric observations of SN~2008D, continued}\\
\hline
\hline
Date       & Telescope+          & Setup        & Seeing   & V  \\
(2008 UT)  & Instrument          &              & (arcsec) &     \\
\hline
01.160  Feb  & LT+RATCam         & UBVRI        &  1.3    & $ 17.43\pm 0.03$\\
05.111  Feb  & REM+ROS+REMIR     & RJHK         &  3.2    & --              \\ 
06.114  Feb  & LT+RATCam         &  BVRI        &  1.3    & $ 17.58 \pm 0.04$ \\
08.162  Feb  & LT+RATCam         & UBVRI        &  2.1    & $ 17.74 \pm 0.03$ \\
11.091  Feb  & TNG+DOLORES       & UBVRI        &  1.8    & $ 18.05 \pm 0.04$\\
13.091  Feb  & REM+ROS+REMIR     & JK           &  3.2    & --              \\ 
\hline
\end{tabular}
\end{center}
\end{table}

%%%%%%%%%%%%%%%%%%%%%%%%%%%%%%%%%%%%%%%%%%%%%%%%%%%%%%%%%%%%%%%%%%%%%%%%%%%%%%

\begin{table}
\begin{center}
\setlength{\tabcolsep}{4pt}
%\resizebox{!}{5cm}{\begin{tabular}{cccccc}
\begin{tabular}{ccccc}
\multicolumn{5}{c}{\bf Supplementary Table S2: 
Spectroscopic observations of SN~2008D}\\
\hline
\hline
Date       & Telescope+           & Setup        & Resolution &  standard  \\
(2008 UT)  & Instrument           &              &   (\AA)    &            \\
\hline
11.097  Jan  & TNG+DOLORES        & LR-B         & 12         & HD93521    \\  
12.045  Jan  & NOT+ALFOSC         & gm4          & 14         & Feige34    \\
13.056  Jan  & TNG+NICS           & IJ           & 7          & Hip10559   \\
13.187  Jan  & TNG+DOLORES        & LR-R         & 13         & Feige34    \\
13.392  Jan  & VLT+FORS2          & 300V; 300I   & 10         & HILT600    \\
14.194  Jan  & NOT+ALFOSC         & gm4          & 14         & G191-B2B   \\
14.944  Jan  & TNG+DOLORES        & LR-B         & 13         & Feige34    \\
16.280  Jan  & VLT+FORS2          & 300V; 300I   & 10         & LTT3864    \\  
25.212  Jan  & VLT+FORS2          & 300V-300I    & 10         & HILT600    \\
28.033  Jan  & CA+CAFOSC          & b200         & 13         & Feige34    \\
28.270  Jan  & NTT+EMMI           & gm2          & 12         & Feige34    \\
28.935  Jan  & TNG+NICS           & IJ+HK        & 7          & Hip10559   \\
28.966  Jan  & As1.82m+AFOSC      & gm4+gm2      & 25         & Feige34    \\
29.958  Jan  & As1.82m+AFOSC      & gm4+gm2      & 25         & Feige34    \\
31.957  Jan  & TNG+NICS           & IJ+HK        & 7          & Hip10559   \\
01.002  Jan  & WHT+ISIS           & R300B+R158R  & 10         & HD93521    \\     
04.107  Feb  & TNG+DOLORES        & LR-B+LR-R    & 12         & Feige34    \\
11.100  Feb  & TNG+DOLORES        & LR-B+LR-R    & 11         & G191-B2B   \\
12.185  Feb  & CA+CAFOSC          & b200         & 13         & Feige34    \\
\hline
\end{tabular}
\end{center}
\end{table}
%%%%%%%%%%%%%%%%%%%%%%%%%%%%%%%%%%%%%%%%%%%%%%%%%%%%%%%%%%%%%%%%%%%%%%%%%%%%%%

\clearpage

\begin{figure}
  \begin{center}
    \includegraphics[width=10cm,angle=270]{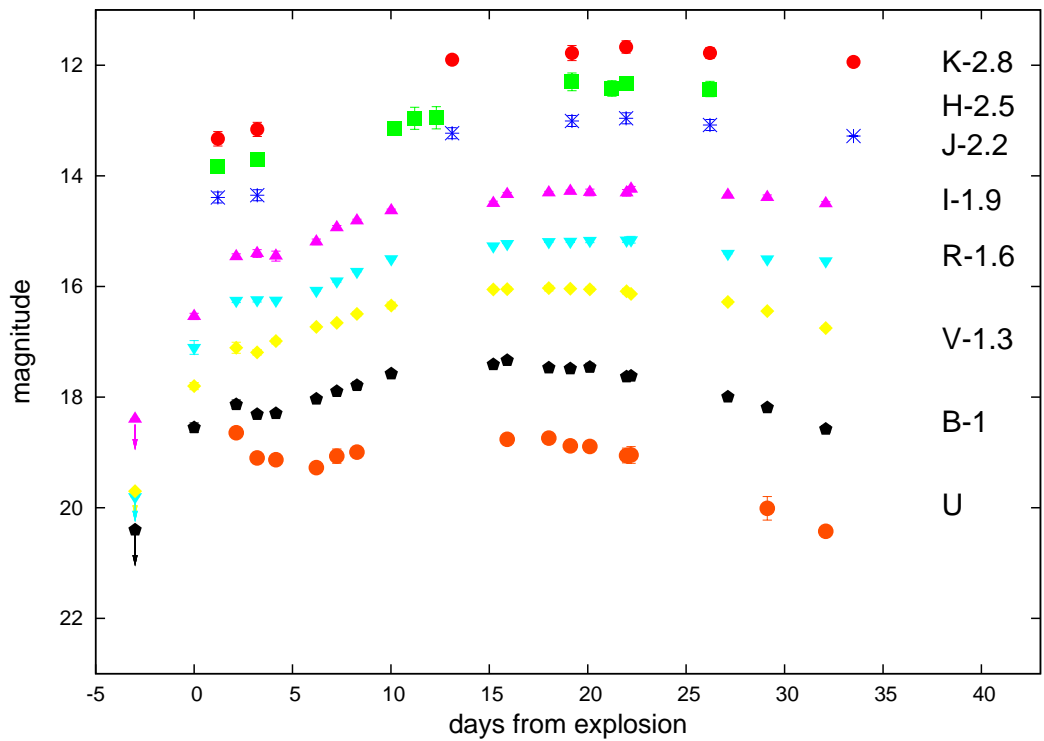}
  \end{center}
\caption{The monochromatic light curves of SN\,2008D.}
  \label{fig:figSOM1}
\end{figure}

\clearpage
%%%%%%%%%%%%%%%%%%%%%%%%%%%%%%%%%%%%%%%%%%%%%%%%%%%%%%%%%%%%%%%%%%%%%%%%%%%%%%

% ----------------------  2: HOST  ---------------------

{\bf 2.~~~The properties of NGC 2770, the host of SN\,2008D/XRF080109.}

\bigskip

NGC 2770 is an Sc galaxy of luminosity class III-IV, similar to the Milky Way
(S1), and very different from the hosts of long GRBs(S2). Its nucleus is rather
diffuse, and according to FIRST radio maps it does not coincide with any radio
peak. The SDSS spectrum of the nucleus is typical of star-forming nuclei of
late-type spiral galaxies: relatively narrow lines (FWHM $<200$\,km\,s$^{-1}$)
and diagnostic emission line ratios are consistent with photoionization by hot
stars. A ratio [\OIII]$\lambda$5007/H$\beta = 0.68$, uncorrected because of the
clearly visible H$\beta$ absorption, makes NGC 2770 a galaxy with an \HII\
nuclear spectrum(S3). Since the Balmer lines are mainly due to photoionization
by hot stars, the star formation rate can be estimated from indicators such as
the H$\alpha$ flux ($0.2 \Msun$\,yr$^{-1}$), IRAS fluxes ($0.45
\Msun$\,yr$^{-1}$), or the radio luminosity at 1.4\,GHz ($0.7
\Msun$\,yr$^{-1}$). From the FIR Luminosity we derive(S4) a SN rate $\sim
0.01$\,SNe\,yr$^{-1}$. These values  indicate that NGC 2770 does not have a very
high star formation luminosity. There is no strong evidence of a hidden
non-thermal source, even at radio frequencies, as already shown by ref. S5, who
failed to find a high surface brightness nuclear radio source in NGC 2770.

\clearpage

% ----------------------  3: REDDENING  ---------------------

{\bf 3.~~~Estimate of the reddening to SN\,2008D and derivation of the
bolometric light curve.}

\bigskip

We estimate dust extinction towards SN 2008D from spectral comparison with other
SNe Ibc.  Using $E(B-V) = 0.16$ for SN\,1983N (S6), an extinction $E(B-V)_{\rm
tot} = 0.65$\,mag for SN\,2008D makes the two SNe almost identical (Fig. 3 in
the main paper).  This value is confirmed by two independent checks: 1) it is
compatible with the neutral hydrogen column density $6 \cdot 10^{21}$\,cm$^{-2}$
estimated from X-ray spectral fits assuming a Galactic gas-to-dust ratio (7); 2)
it yields optical and infrared colour light curves similar to those of other SNe
Ib/c. Considering the uncertainties, we adopt for SN\,2008D $E(B-V)_{tot}= 0.65
\pm 0.15$. Most of the extinction occurs in the host galaxy, since Galactic
extinction is very weak [$E(B-V)_{\rm Gal}$ = 0.02 mag; ref. 8].  

We computed the bolometric light curve using the data in Table 1 together with
the measurements reported in refs. 9 and 10.

\clearpage

% ----------------------  4: X-RAY DATA ---------------------

{\bf 4.~~~X-ray data and models}

\bigskip

\noindent

The XRT light curve (Figure S2, top panel) was fitted with a function ${\rm
Count Rate} = N(t-t_0) \sim e^{-\alpha (t-t_0)}$, where $N = 14.5 \pm 2.2$ and
$\alpha = 1.09 \pm 0.13$, yielding $\chi^2/{\rm d.o.f.} = 13.1/16$.  From this
fit, the onset of the XRF can be estimated to have occurred on $JD =
2454475.06413$. The X-ray flux reaches maximum in $\approx 92$\,s, with an
uncertainty of $\sim 30$\,s. The decaying part of the X-ray curve can also be
fitted with a power-law with index $\alpha = 1.52 \pm 0.16$. The number of
photons is too small to perform time-resolved spectroscopy. Instead, we show the
time evolution of the hardness ratio $HR = CR(1.5 - 10\,{\rm keV})/CR(0.3 -
1.5\,{\rm keV}$ (Fig. S2, bottom panel).  The spectrum hardens during the rise
to maximum and softens afterwards, a typical behaviour of X-ray flares in GRBs
and in FRED profiles.

The source is affected by pile-up only within a 3-pixel radius. Therefore we
extracted the photons from an annular region with a 3-pixel inner radius and  
a 30-pixel outer radius, yielding a total of 539 photons.

The spectrum can be fitted with a simple power-law typical of GRBs, with photon
index $\Gamma = 2.29^{+0.26}_{-0.24}$ and a host galaxy hydrogen column density
of $N_H = 6.42^{+1.42}_{-1.18} \cdot 10^{21}$\,cm$^{-2}$ ($\chi^2/{\rm d.o.f.} =
21.1/23$).  Alternatively, a model with a black body with temperature $kT =
0.20^{+0.30}_{-0.20}$\,keV superimposed on a power-law with photon index $\Gamma
= 2.21^{+0.48}_{-0.93}$, and a host galaxy hydrogen column density $N_H =
6.96^{+7.03}_{-2.71} \cdot 10^{21}$\,cm$^{-2}$ yields $\chi^2/{\rm d.o.f.} =
20.6/21$. The total intensity is $1.85 \times 10^{-10}$\, \ergcms, and
the black body intensity is no more than 14\% of it, thus
the presence of a black body remains uncertain (11).

The X-ray luminosity can be computed using the conversion from count rate to
flux 1\,count\,s$^{-1}$ $= 1.17 \cdot 10^{-10}$\,\ergcms, derived from the
spectral power-law fit corrected for $N_H$ absorption. Over the 605 sec of the
X-ray event, after applying a PSF correction factor of 1.88, this yields a
fluence of $1.1 \cdot 10^{-7}$\, erg\,cm$^{-2}$, corresponding to a total energy
$1.3 \cdot 10^{46}$\,erg. The peak luminosity is $\sim 8 \cdot
10^{43}$\,\ergsec. Assuming that the black-body component is a constant fraction
(14\%) of the total luminosity at any time, its luminosity, combined with the
high black-body temperature $(3.8 \cdot 10^6$\,K), implies an emitting radius
$R_{{\rm ph}} \sim 10^{10}$\,cm, an order of magnitude smaller than the size of
Wolf-Rayet stars.

%%%%%%%%%%%%%%%%%%%%%%%%%%%%%%%%%%%%%%%%%%%%%%%%%%%%%%%%%%%%%%%%%%%%%%%%%%%%%%

\begin{figure}
 \begin{center}
 \includegraphics[width=14cm,angle=0]{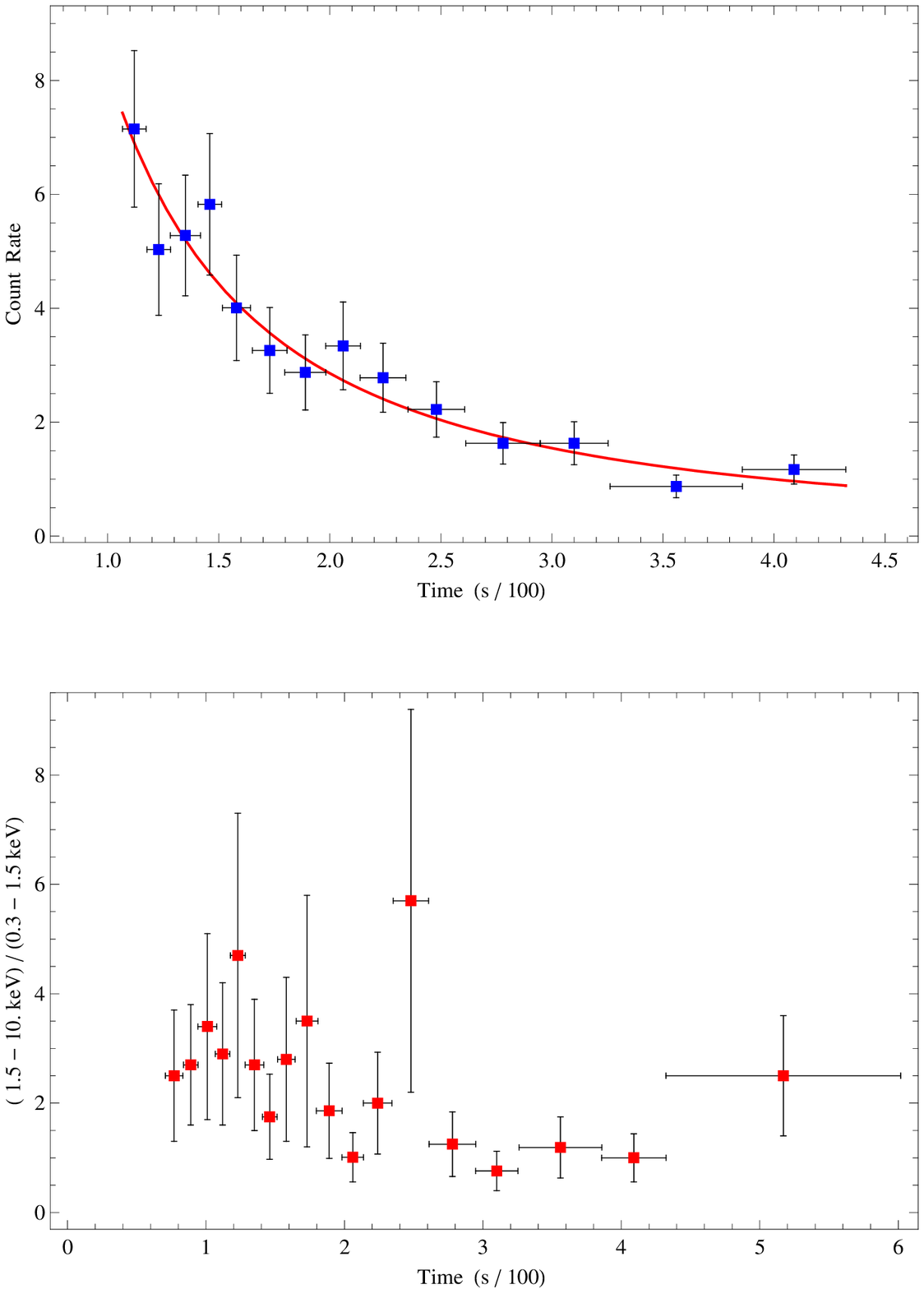}
 \end{center}
 \caption{Top: the decaying part of the X-ray curve and a fit with a 
 power-law with index $\alpha = 1.52 \pm 0.16$;
 Bottom: time evolution of the hardness ratio 
 $HR = CR(1.5 - 10\, {\rm keV})/CR(0.3 - 1.5\, {\rm keV})$.}
 \label{x-figure}
\end{figure}

\clearpage

% ----------------------  5: SPECTRAL MODELS  ---------------------

{\bf 5.~~~Spectral modelling.}

\bigskip

To estimate the physical parameters that describe the supernova explosion we
construct a series of radiative transfer models to derive synthetic spectra. As
time elapses after the explosion of the supernova the ejecta expand and become
diluted, progressively exposing deeper layers. Modelling the spectra as they
evolve enables us to infer the structure of the ejected material. Among the
parameters that can be determined in this way are the luminosity, the postion of
the momentary photosphere, and the composition and velocity of the line-forming
layers of the ejecta at each epoch.

For the spectral models we use a Monte Carlo radiative transfer code (S12, S13,
S14). The code employs an approximate description of non-LTE suitable for the
analysis of supernova spectra during the photospheric epochs while retaining
physically meaningful relationships between model parameters.  Here we use a
version of the code that uses a depth-dependent composition structure (S15). For
all models in the series a density structure $\rho(v)$ is adopted (for SNe a
Hubble-like expansion law $r=v \cdot t$ allows us to use the ejecta velocity as
a time-independent radial variable). Input parameters for each individual model
are the luminosity at the given epoch, a lower boundary velocity, and the
composition above this velocity. This lower boundary represents the
pseudo-photosphere from which all radiation is assumed to emerge, which is a
good approximation at early times.  Lacking self-consistent explosion models we
use a parameterized density structure which we constrain iteratively with the
help of light curve and spectral models. Our model does not take into account
non-thermal excitation and ionization by fast particles from the radioactive
decay of \Nifs. This is a fair approximation for most elements but it fails for
helium. Because of the high ionization potential of \HeI, non-thermal processes
are the main contribution to the excitation of this elements.  Therefore, we
cannot self-consistently derive the opacity of \HeI\ lines, which would anyway
require a detailed model of the distribution of \Nifs\ and of the geometry of
the ejecta.

The very broad line features at early epochs, up to day 5 after explosion,
indicate the presence of appreciable amounts of matter at high velocities above
$v \sim 30,000$\,\kms.  Starting approximately 2008 January 15, ($t=6\,$d) the
lines become narrower and the blue-shifted high-velocity components disappears.
This suggests a significant steepening of the density gradient at lower
velocities. The last spectrum considered here was obtained on 2008 February 11
($t=33\,$d). For this model we use a lower boundary velocity of 7500\,\kms. The
line widths of this last spectrum suggest that the density gradient flattens
below a velocity of about 9000\,\kms. Deeper layers are not yet accessible to
observations.  Fig. S3 shows the time evolution of the photospheric velocity. 

A rough estimate of the parameters is obtained from the similarity of SN~2008D
to SNe\,2002ap and 2006aj at the earliest times and to SN\,1994I at later
phases. The models of these objects (S16, S17, and S18) are used as a
guideline. The evolution of synthetic spectra is shown in Fig. S4.

For the models shown here we use a broken power-law with a power-law index of
$n=-2.0$ in the region inside of 9000\,\kms, $n=-7.5$ between 9000\,\kms\ and
17,000\,\kms, and $n=-5.5$ above 17,000\,\kms.  We set the absolute density to
match the observed spectral features, and obtain a total ejected mass of $\sim 7
\msun$ with a kinetic energy of $\sim 6 \times 10^{51}$\,erg, of which a mass
$\sim 0.03 \msun$ with a \KE$\sim 5 \times 10^{50}$\,erg is located at $v >
30,000$\,\kms.

The first spectrum we consider here was taken on 12 January, 3 days after the
explosion.  An interesting feature of the spectra is the absence of a strong
\OI\ absorption at $\sim\,7300\,${\AA}. The \OI\ lines at 7774\,\AA\ that are
normally responsible for this feature are very strong. Therefore we can only
accommodate small amounts of oxygen in the outer layers of the ejecta. At early
epochs oxygen could be so highly ionized that \OI\ lines disappear, but the
presence of \CaII\ and \FeII\ in the spectra makes this hypothesis unlikely
because those species should not be present in an environment where oxygen is
mostly ionized.

Overall, we can model the spectra with only a mild variation of the composition
with depth. Spectra before maximum exhibit broad but shallow features of Ca, Si,
and Fe-group elements suggesting the presence of only small amounts of absorbing
material.  In the outer part of the ejecta above $v\sim 14,000$\,\kms\ we assume
a composition dominated by He. The attenuation of the flux in the blue requires
the presence of some Ti and other Fe-group elements that do not produce strong
individual line features but block radiation in the blue and UV through a large
number of overlapping weaker lines.

At epochs around and after maximum the spectra show distinctly narrower line
features that can be attributed the \HeI, \OI, \CaII\ and \FeII. There is also
some indication for \CI\ in the red and infrared region (S19). In the inner
part we assume a composition where C and O dominate with a slightly enhanced
contribution from heavy elements, likely including some decay products of
radioactive \Nifs.

%%%%%%%%%%%%%%%%%%%%%%%%%%%%%%%%%%%%%%%%%%%%%%%%%%%%%%%%%%%%%%%%%%%%%%%%%%%%%%

\begin{figure}
  \begin{center}
    \includegraphics[width=14cm]{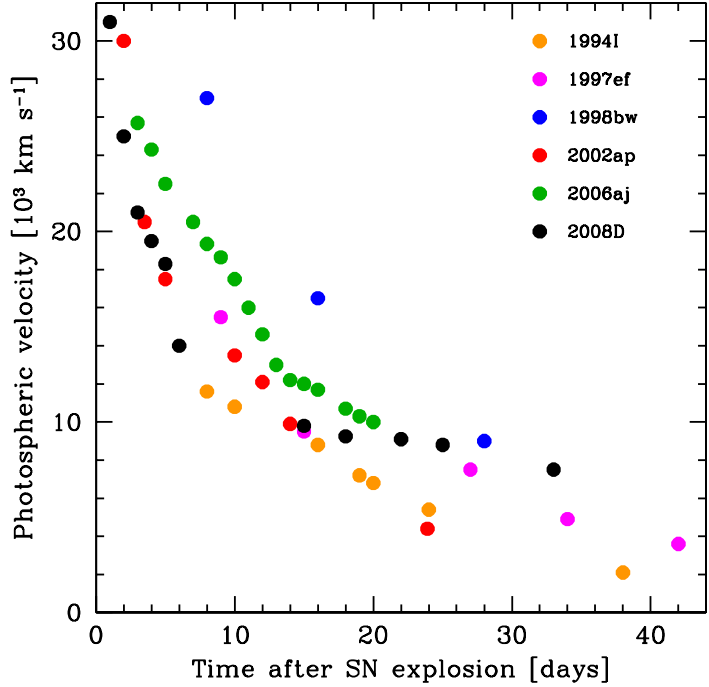}
  \end{center}
%\centerline{\psfig{file=SI_Fig1_velevol.eps,width=8cm}}
\caption{The temporal evolution of the lower boundary (photosphere) velocity 
used in the models for SN\,2008D compared to the cases of other SNe\,Ic.
SN\,2008D starts out with a very high velocity, like SN\,2002ap, then 
transitions to lower velocities like SN\,1994I. This is the phase when broad
lines disappear. At later times the evolution is slower than that of SN\,1994I,
indicating a large mass with a small density gradient in the inner layers,
similar to SN\,1997ef. Only SNe with velocities larger than SNe\,2008D or 
2002ap were accompanied by a GRB or an XRF. 
}
  \label{fig:figSOM3}
\end{figure}

%%%%%%%%%%%%%%%%%%%%%%%%%%%%%%%%%%%%%%%%%%%%%%%%%%%%%%%%%%%%%%%%%%%%%%%%%%%%%%

\begin{figure}[p]
  \begin{center}
    \includegraphics[width=12cm]{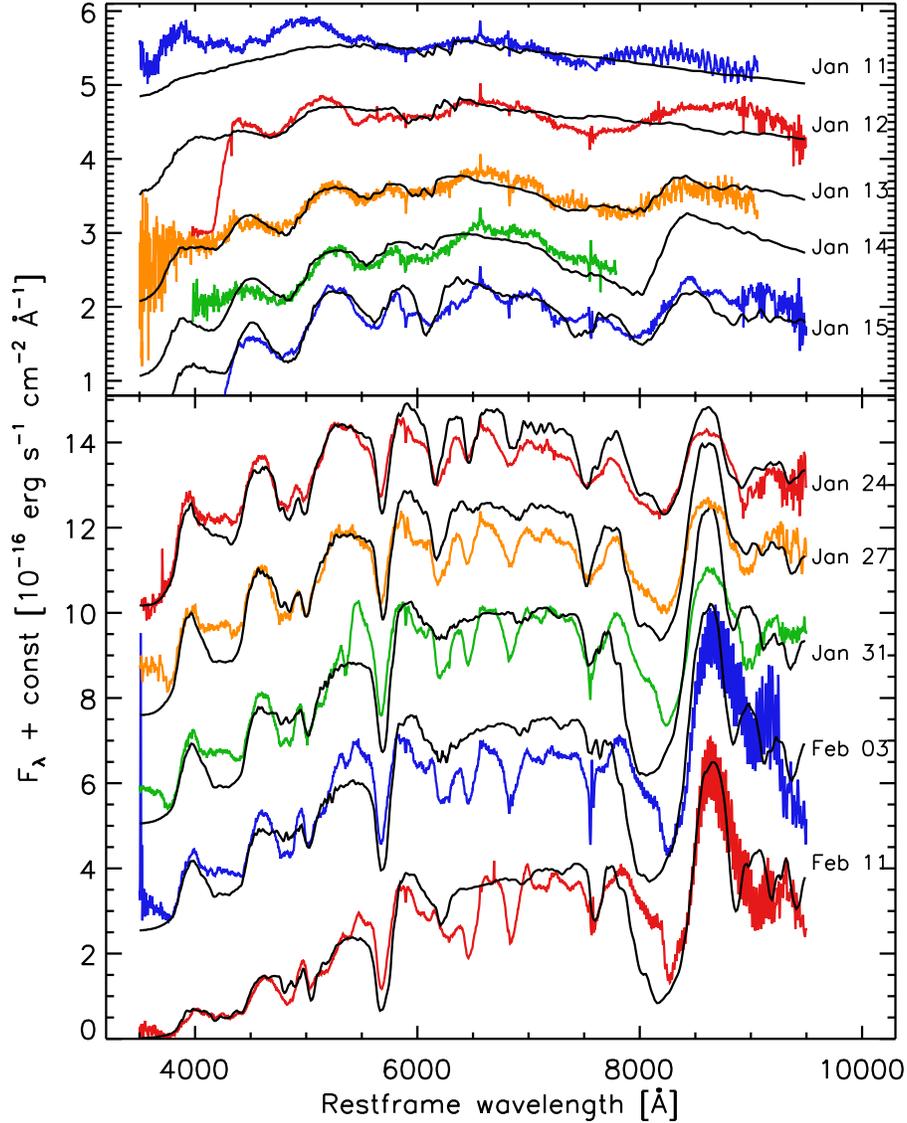}
  \end{center}
  \caption{Series of model spectra for the sequence of early time spectra of
  SN~2008D.   
The optical depth in \HeI\ lines in the model of 24 January has been enhanced to
mimic the effect of non-thermal excitations by fast electrons generated in the
decay of $^{56}$Ni. While we cannot constrain the abundance of He this way this
method allows us to identify He lines in the spectrum. The first two spectra are
contaminated by the emission of the afterglow which is not described by the
model. Therefore, the luminosity needed to match those spectra is too high to
give a consistent description leading to an over-ionization of most species.
This affects in particularly the \CaII\ IR triplet near 8000\,\AA, which is not 
reproduced by the models in the first two epochs. 
}
  \label{fig:spectra}
\end{figure}

\clearpage

% ----------------------   REFS  ---------------------

{\bf References.}

\bigskip

\begin{enumerate}

\item  van den Bergh, S., Li, W., \& Filippenko, A.~V.\, 
% Classifications of the Host Galaxies of Supernovae, Set II  
 \pasp, 115, 1280 (2003).

\item  Fruchter, A., et al.\ , 
% Long gamma-ray bursts and core-collapse supernovae have different environments.
 \nat 441, 463-468 (2006).

\item  Veilleux, S., \& Osterbrock, D.~E.\ , 
% Spectral classification of emission-line galaxies,  
\apjs, 63, 295 (1987).
 
\item  Mannucci, F., et al.\ , 
% The infrared supernova rate in starburst galaxies, 
\aap, 401, 519 (2003).

\item  Ulvestad, J.~S., \& Ho, L.~C. 
% A Search for Active Galactic Nuclei in Sc Galaxies with H II Spectra, 
 \apj, 581, 925 (2002).

\item Clocchiatti, A., \& Wheeler, J.C. 
% On the Light Curves of Stripped-Envelope Supernovae, 
\apj, 491, 375 (1997).

\item Page, K.L., {\it et al.} 
% Refined Swift-XRT analysis of the transient in NGC 2770, 
 GCN 7164 (2008).
 
\item  Schlegel, D.J., Finkbeiner, D.P., \& Davis, M., 
% Maps of Dust Infrared Emission for Use in Estimation of Reddening and 
% Cosmic Microwave Background Radiation Foregrounds,  
 \apj, 500, 525 (1998).

\item  Immler, S., et al.\ 
 GRB Coordinates Network, 7168, 1 (2008).

\item  Li, W., Chornock, R., Foley, R.~J., Filippenko, A.~V., Modjaz, M.,
 Poznanski, D., \& Bloom, J.~S.\, GRB Coordinates Network, 7176, 1  (2008).

\item Li, L.-X.,  astro-ph 0803.0079 (2008).

\item  Mazzali, P.~A. \& Lucy, L.~B., 
% The application of Monte Carlo methods
%  to the synthesis of early-time supernovae spectra, 
  \aap, 279, 447 (1993).

\item  Lucy, L.~B., 
% Improved Monte Carlo techniques for the spectral synthesis of supernovae, 
 \aap, 345, 211 (1999).

\item  Mazzali, P.~A., 
% Applications of an improved Monte Carlo code to the synthesis of 
% early-time Supernova spectra, 
 \aap, 363, 705 (2000).

\item  Stehle, M., Mazzali, P.~A., Benetti, S., \& Hillebrandt, W., 
% Abundance stratification in Type Ia supernovae - I. The case of SN 2002bo,
  \mnras, 360, 1231 (2005).

\item  Mazzali, P.~A., et~al.,  
% The Type Ic Hypernova SN 2002ap, 
 \apjl, 572, L61 (2002).

\item  Mazzali, P.~A., et~al., 
% A neutron-star-driven X-ray flash associated
%  with supernova SN 2006aj, 
\nat, 442, 1018 (2006).

\item  Sauer, D.~N., et~al., 
% The properties of the `standard' Type Ic
%  supernova 1994I from spectral models, 
  \mnras, 369, 1939 (2006).
  
\item  Valenti, S., et~al., 
% The Carbon-rich Type Ic SN 2007gr: The Photospheric Phase, 
 \apjl, 673, L155 (2008).

\end{enumerate}


\begin{thebibliography}{30}

\bibitem[1]{sod08a}  {Berger}, E., \& Soderberg, A.M.,
%Bright X-ray Transient in NGC 2770 - A low-luminosity XRF?
GCN 7159 (2008).

\bibitem[2]{deng08} {Deng}, J., \& Zhu, Y.,
%Bright X-ray Transient (a XRF?) in NGC 2770 - A SN optical counterpart?
GCN 7160 (2008).

\bibitem[3]{val7171} {Valenti}, S., {\it et al.},
%Spectroscopic classification of SN 2008D (Transient in NGC 2770).
GCN 7171 (2008).

\bibitem[4]{fil97} {Filippenko}, A.V.,
%Optical Spectra of Supernovae.
{\it Ann. Rev. Astron. Astrophys.} {\bf 35}, 309 (1997).

\bibitem[5]{gal98}  {Galama}, T.~J., {\it et al.},
%An unusual supernova in the error box of the gamma-ray burst of 25 April 1998.
{\it Nature} {\bf 395}, 670 (1998).

\bibitem[6]{iwa98} {Iwamoto}, K., {\it et al.},
%A hypernova model for the supernova associated with the gamma-ray burst
%of 25 April 1998.
{\it Nature} {\bf 395}, 672 (1998).

%\bibitem[{Stanek} {\it et al.}<3>]{3-sta03} {Stanek}, K.Z., {\it et al.} 2003,
%Spectroscopic Discovery of the Supernova 2003dh Associated with GRB 030329.
%{\it Astrophys. J.} {\bf 591}, L17-L20 (2003)

%\bibitem[{Mazzali} {\it et~al.}<4>]{m03}  {Mazzali}, P. A., {\it et al.}
%The Type Ic Hypernova SN 2003dh/GRB 030329.
%{\it Astrophys. J.} {\bf 599}, L95-L98 (2003).

\bibitem[7]{pian06}  {Pian}, E., {\it et al.},
%An optical supernova associated with the X-ray flash XRF 060218.
{\it Nature}  {\bf 442}, 1011 (2006).

\bibitem[8]{m06aj}  {Mazzali}, P. A., {\it et al.},
%A neutron-star-driven X-ray flash associated with supernova SN 2006aj.
{\it Nature}  {\bf 442}, 1018 (2006).

\bibitem[9]{m02ap} {Mazzali}, P. A., {\it et al.},
%The Type Ic Hypernova SN 2002ap
{\it Astrophys. J.} {\bf 572}, L61 (2002).

%\bibitem[{Valenti} {\it et~al.}<7>]{val7163} {Valenti}, S., Turatto, M.,
%Navasardyan, H., Benetti, S., \&  Cappellaro, E.
%Early OT detection of XRF in NGC 2770 in Asiago frames
%GCN 7163 (2008)

%\bibitem[9]{woobl06} {Woosley}, S.E., \&  Bloom, J.S.,
%The Supernova Gamma-Ray Burst Connection.
%{\it Ann. Rev. Astron. Astrophys.} {\bf 44}, 507 (2006).

%\bibitem[{Kong} {\it et~al.}<3>]{kong08} {Kong}, A.K.H., Soderberg, A.M.,
%Berger, E., Rea, N., \& Maccarone, T.
%Swift UVOT Observations of the X-ray transient in NGC 2770.
%&GCN 7170 (2008)

\bibitem[10]{fru06} {Fruchter}, A.S., {\it et al.},
%Long gamma-ray bursts and core-collapse supernovae have different environments.
{\it Nature} {\bf 441}, 463 (2006).

%\bibitem[{Turatto}<14>]{tur03} {Turatto}, M., Benetti, S., \& Cappellaro, E.
%Variety in Supernovae.
%{\it From Twilight to Highlight: The Physics of Supernovae}, {} 200-209

%\bibitem[{Schlegel}<15>]{schl98} {Schlegel}, D.J., Finkbeiner, D.P., \& Davis, M.
%Maps of dust infrared extinction for use in estimation of reddening and cosmic
%microwave background radiation foregrounds.
%{\it Astrophys. J.} {\bf 500}, 525-553

%\bibitem[{Page} {\it et~al.}<16>]{page08} {Page}, K.L., {\it et al.}
%Refined Swift-XRT analysis of the transient in NGC 2770.
%GCN 7164 (2008)

\bibitem[11]{str02} {Stritzinger}, M., {\it et al.},
%Optical Photometry of the Type Ia Supernova 1999ee and the
%Type Ib/c Supernova 1999ex in IC 5179.
{\it Astron. J.} {\bf 124}, 2100 (2002).

%\bibitem[12]{hamuy02} {Hamuy}, M., {\it et al.},
%Optical and Infrared Spectroscopy of SN 1999ee and SN 1999ex.
%{\it Astron. J.} {\bf 124}, 417 (2000).

\bibitem[12]{mod7212} {Modjaz}, M., Chornock, R., Foley,
R.J., Filippenko, A.V., Li W., \& Stringfellow G.,
%Transient 080109/SN 2008D: Spectroscopic Evolution and Re-Classification.
GCN 7212 (2008).

\bibitem[13]{m00} {Mazzali}, P. A., Iwamoto, K., Nomoto, K.,
%A Spectroscopic Analysis of the Energetic Type Ic Hypernova SN 1997ef.
{\it Astrophys. J.} {\bf 545}, 407 (2002).

\bibitem[14]{tom05} {Tominaga}, N., {\it et al.},
%The Unique Type Ib Supernova 2005bf:
%A WN Star Explosion Model for Peculiar Light Curves and Spectra.
{\it Astrophys. J.} {\bf 633}, L97 (2005).

\bibitem[15]{ml98} {Mazzali}, P. A., \& Lucy, L.B.,
%The 1.05-$\mu$m feature in the spectrum of the Type Ia supernova 1994D:
%He in SNe Ia?
{\it Mon. Not. R. Astron. Soc.} {\bf 295}, 428 (1998).

\bibitem[16]{lucy91} {Lucy}, L.B.,
%Nonthermal excitation of helium in type Ib supernovae.
{\it Astrophys. J.} {\bf 383}, 308 (1991).

\bibitem[17]{sauer06} {Sauer}, D., {\it et al.},
%The properties of the `standard' Type Ic supernova 1994I from spectral models
{\it Mon. Not. R. Astron. Soc.} {\bf 369}, 1939 (2006).

\bibitem[18]{maeda03} {Maeda}, K., {\it et al.},
{\it Astrophys. J.} {\bf 593}, 931 (2003).

\bibitem[19]{maeda06} {Maeda}, K., {Mazzali}, P.A.,, \& {Nomoto}, K.,
%Optical Emission from Aspherical Supernovae and the Hypernova SN 1998bw
{\it Astrophys. J.} {\bf 645}, 1331 (2006).

\bibitem[20]{McF&W99}  {MacFadyen}, A. E., \& {Woosley}, S. E.,
%Collapsars: Gamma-Ray Bursts and Explosions in ``Failed Supernovae''.
{\it Astrophys. J.} {\bf 524}, 262 (1999).

%\bibitem[20]{gmdv07} {Guetta}, D., Della Valle, M.,
%On the Rates of Gamma-Ray Bursts and Type Ib/c Supernovae
%{\it Astrophys. J.} {\bf 657}, L73 (2007).

\bibitem[21]{clocch97} {Clocchiatti}, A., \& Wheeler, J.C.,
%On the Light Curves of Stripped-Envelope Supernovae.
{\it Astrophys. J.} {\bf 491}, 375 (1997).

\bibitem[22]{capp99} {Cappellaro}, E., {Evans}, R., {Turatto}, M.,
%A new determination of supernova rates and a comparison with indicators
%for galactic star formation
{\it A\&A} {\bf 351}, 459 (1999).

\bibitem[23]{sod08} {Soderberg}, A.M., {\it et al.}, Nature 453, 469 (2008).

\bibitem[24]{camp06} {Campana}, S., {\it et al.},
%The association of GRB060218 with a supernova and the evolution of the shock
%wave.
{\it Nature} {\bf 442}, 1008 (2006).

%\bibitem[24]{li08} {Li}, L.-X.,  astro-ph 0803.0079 (2008).

\bibitem[25]{m05} {Mazzali}, P.A., {\it et al.},
%An Asymmetric Energetic Type Ic Supernova Viewed Off-Axis,
%and a Link to Gamma Ray Bursts
{\it Science} {\bf 308}, 1284 (2005).

\bibitem[26]{mae08} {Maeda}, K., et al.,
%Asphericity in Supernova explosions from late-time spectroscopy.
{\it Science} {\bf 319}, 1220 (2008)

\bibitem[27]{arn82}  {Arnett}, W. D.,
%Type I supernovae. I - Analytic solutions for the early part of the light curve.
{\it Astrophys. J.} {\bf 253}, 785 (1982).

\bibitem[28]{m03lw} {Mazzali}, P.A., {\it et al.},
%Models for the Type Ic Hypernova SN 2003lw associated with GRB 031203.
{\it Astrophys. J.} {\bf 645}, 1323 (2006).

\bibitem[29]{capp97} {Cappellaro}, E., {\it et al.},
%SN IA light curves and radioactive decay
{\it Astron. Astrophys.} {\bf 328}, 203 (1997).

%\bibitem[{Folatelli}<14>]{fol06} {Folatelli}, G., {\it et al.}
%SN 2005bf: A Possible Transition Event between Type Ib/c Supernovae.
%and Gamma-Ray Bursts
%{\it Astrophys. J.} {\bf 641}, 1039-1050

%\bibitem[{Maeda} {\it et~al.}<15>]{mae07} {Maeda}, K., {\it et al.}
%The Unique Type Ib Supernova 2005bf at Nebular Phases:
%A Possible Birth Event of a Strongly Magnetized Neutron Star.
%{\it Astrophys. J.} {\bf 666}, 1069-1082 (2007)

%\bibitem[{Valenti} {\it et~al.}<26>]{val7221} {Valenti}, S., D'Elia, V., Della
%Valle, M., Benetti, S., Chincarini, G., Mazzali, P.A., \& Antonelli, L.A.
%Transient 080109/SN 2008D: Spectroscopic Evolution.
%GCN 7221 (2008)

%\bibitem[27]{tho04} Thompson, T.A., Chang, P., \& Quataert, E.,
%Magnetar Spin-Down, Hyperenergetic Supernovae, and Gamma-Ray Bursts.
%{\it Astrophys. J.} {\bf 611}, 380 (2004).

%\bibitem[{Waxman}<27>]{wax04}  {Waxman}, E.,
%The Nature of GRB 980425 and the Search for Off-Axis Gamma-Ray Burst Signatures
%in Nearby Type Ib/c Supernova Emission
%{\it Astrophys. J.} {\bf 602}, 886-891 (1999).

%\bibitem[{Soderberg} {\it et~al.}<1>]{sod08}  {Soderberg}, A.M., Berger, E.,
%Fox, D., Cucchiara, A.,  Rau, A., Ofek, E., Kasliwal, M., \& Cenko, S.B.
%Optical Spectroscopy of the Transient in NGC 2770.
%GCN  7165 (2008)

\end{thebibliography}
\end{document}